# Digital twin-assisted three-dimensional electrical capacitance tomography for multiphase flow imaging

Shengnan Wang, Yi Li, Zhou Chen, *Student Member, IEEE*, and Yunjie Yang, *Member, IEEE*

***Abstract*—Three-dimensional electrical capacitance tomography (3D-ECT) has shown promise for visualizing industrial multiphase flows. However, existing 3D-ECT approaches suffer from limited imaging resolution and lack assessment metrics, hampering their effectiveness in quantitative multiphase flow imaging. This paper presents a digital twin (DT)-assisted 3D-ECT, aiming to overcome these limitations and enhance our understanding of multiphase flow dynamics. The DT framework incorporates a 3D fluid-electrostatic field coupling model (3D-FECM) that digitally represents the physical 3D-ECT system, which enables us to simulate real multiphase flows and generate a comprehensive virtual multiphase flow 3D imaging dataset. Additionally, the framework includes a deep neural network named 3D deep back projection (3D-DBP), which learns multiphase flow features from the virtual dataset and enables more accurate 3D flow imaging in the real world. The DT-assisted 3D-ECT was validated through virtual and physical experiments, demonstrating superior image quality, noise robustness and computational efficiency compared to conventional 3D-ECT approaches. This research contributes to developing accurate and reliable 3D-ECT techniques and their implementation in multiphase flow systems across various industries.**

## I. INTRODUCTION

MULTIPHASE flows serve a pivotal role in many industrial processes such as renewable energy, oil and gas, chemical and petrochemical sectors, pharmaceutical industry, food and beverage industry, and mining and mineral processing [1]. The accurate visualization and characterization of these flows are crucial for process optimization, safety and efficiency [2], [3]. Electrical capacitance tomography (ECT) is a promising non-invasive technique for imaging industrial multiphase flows [4]. Traditional two-dimensional (2D) ECT, while providing valuable insights, misses information in the vertical dimension, limiting our comprehension of flow dynamics in complex three-dimensional (3D) environments [5]. Recent research has been shifting focus towards advancing 3D electrical capacitance tomography (3D-ECT) [6-9]. By incorporating real-time and 3D imaging capabilities, 3D-ECT provides a more detailed and comprehensive assessment of flow dynamics, phase distribution, and flow patterns [10]. These advancements hold great promise for enhancing process understanding and ensuring the efficient and safe functioning of multiphase flow systems across various industries.

However, quantitatively imaging multiphase flows using existing 3D-ECT approaches faces challenges due to the ill-posed nature of the inverse problem and the complex features of multiphase flows [10], [11]. To address these challenges, it is crucial to 1) develop robust and efficient inversion algorithms tailored to 3D-ECT-based multiphase flow imaging, and 2) assess the image quality against true 3D flow profiles, rather than simplified static phantoms.

Traditional iterative methods, such as Landweber iteration (LI) [12] and Total Variation (TV) [13], have been employed for 3D image reconstruction in electrical tomography (ET). However, these methods often require significant computational time, thus unsuitable for real-time applications. Deep learning (DL) provides potential solutions for the nonlinear 3D-ECT inverse problem, which can possibly achieve high-precision and real-time performance. Several networks including convolutional neural network (CNN) [14] and graph convolutional network (GCN) [6], have been explored for high-quality 3D image reconstruction in ET. However, these networks heavily rely on static phantom data and struggle to transfer to realistic multiphase flow imaging scenarios that involve dynamic flow behaviours. Furthermore, the evaluation of the accuracy and reliability of 3D-ECT for multiphase flow imaging is challenging due to the absence of ground truth data.

Recently, we proposed a virtual evaluation platform for 2D electrical tomography, allowing for the quantitative assessment of 2D image reconstruction algorithms in dynamic multiphase flows [15]. We also reported a DT framework for 2D ECT,

This work was supported in part by the National Natural Science Foundation of China under Grant 51906209 and European Union's Horizon 2020 Research and Innovation Programme under the Marie Sklodowska-Curie actions COFUND Transnational Research and Innovation Network at Edinburgh under Grant 801215.

Shengnan Wang is with the College of Metrology and Measurement Engineering, China Jiliang University, Zhejiang 310018, China. (e-mail: 691656937@qq.com).

Yi Li is with the Tsinghua Shenzhen International Graduate School, Tsinghua University, Shenzhen 518055, China (e-mail: liyi@sz.tsinghua.edu.cn).

Zhou Chen and Yunjie Yang are with the SMART Group, Institute for Digital Communications, School of Engineering, The University of Edinburgh, Edinburgh EH9 3FG, U.K. (e-mail: y.yang@ed.ac.uk).

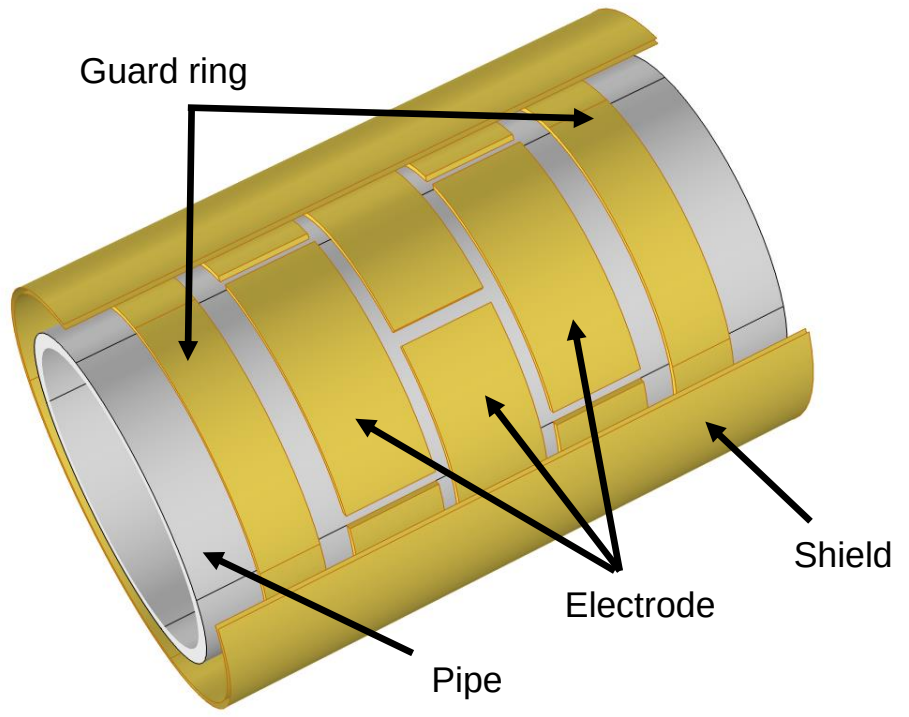

Fig. 1. 3D electrical capacitance tomography sensor.

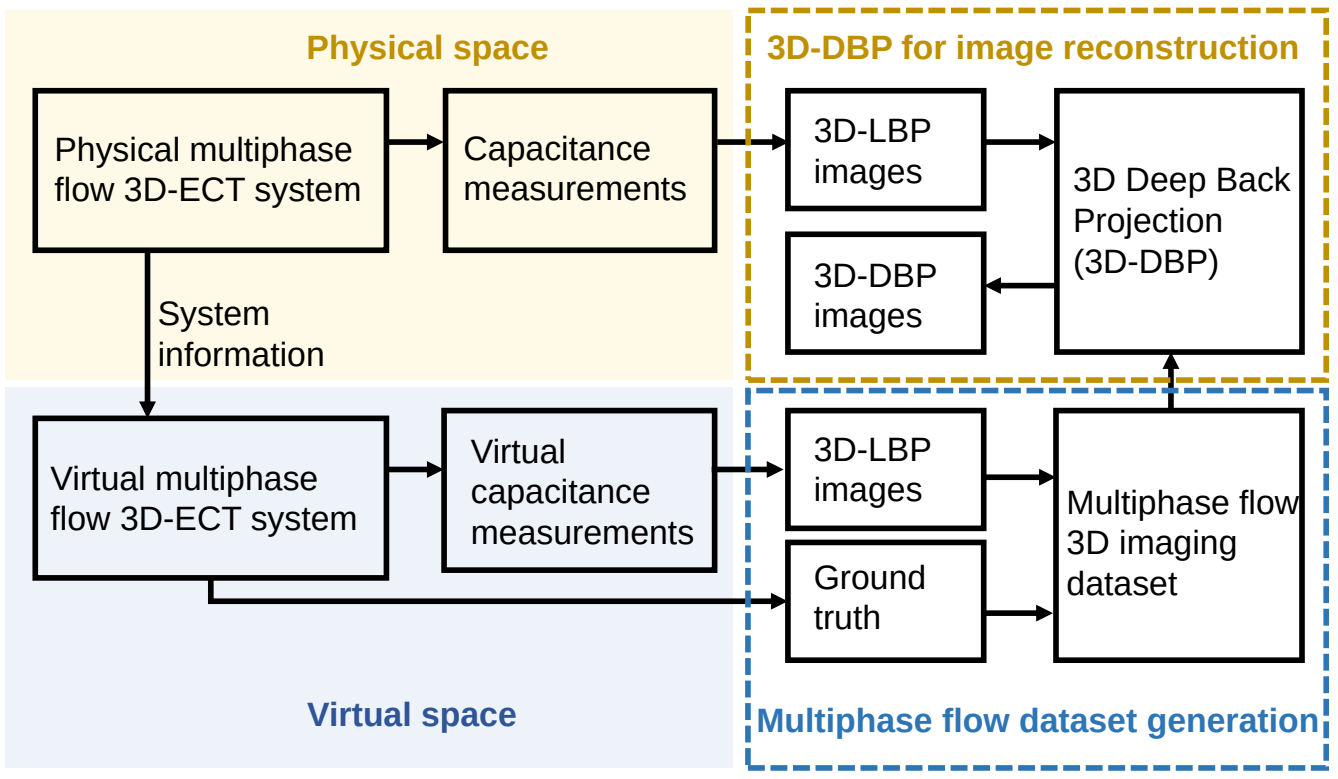

Fig. 2. Flowchart of the digital twin-assisted 3D-ECT.

which demonstrated remarkable enhancements in multiphase flow imaging in the 2D setup [16]. Building upon our previous efforts, this paper further extends the capabilities of our DT framework to encompass 3D multiphase flow imaging. The key contributions include:

1) A 3D fluid-electrostatic field coupling model (3D-FECM) is developed to digitally represent the 3D-ECT system and the multiphase flow pipes, enabling the modelling of the dynamic behaviour of real multiphase flows and generating virtual 3D-ECT measurements;

2) We conduct dynamic coupling simulations and generate a 3D virtual dataset comprising 3D-ECT data and corresponding 3D flow profiles. We propose a deep neural network, 3D deep back projection (3D-DBP), to enable 3D quantitative multiphase flow imaging.

## II. Methodology

### A. Principle of 3D-ECT

Fig. 1 shows our 3D-ECT sensor with three electrode layers, each containing four electrodes, totaling twelve electrodes. The pipe encompassing the sensor has an inner diameter of 50 mm and an outer diameter of 55 mm. The axial length of each electrode is 15 mm, and the electrode coverage angle is 85°. The principle of 3D-ECT is to estimate the spatial distribution of dielectric permittivity $\varepsilon(x,y,z)$ in the domain of interest (DOI) through inter-electrode capacitance measurement $C_m$. The relationship between $\varepsilon(x,y,z)$ and $C_m$ is governed by [17]

$$C_m = \frac{Q}{V} = -\frac{1}{V}\iint_{\Gamma} \varepsilon(x,y,z)\nabla\varphi(x,y,z)d\Gamma \quad (1)$$

where $Q$ is the electric charge on the measured electrode; $V$ is the inter-electrode potential difference; $\Gamma$ is the electrode surface; $(x, y, z)$ denotes the voxel coordinate of the DOI; $\varphi(x, y, z)$ is the potential distribution.

In practice, (1) can be simplified to a linear normalized matrix form:

$$\mathbf{C} = \mathbf{S}\mathbf{g} \quad (2)$$

where $\mathbf{C}=(\mathbf{C_r}-\mathbf{C_l})/(\mathbf{C_h}-\mathbf{C_l})\in\mathbb{R}^M$ signifies the normalized capacitance measurements, wherein vector division '/' denotes element-wise division. $\mathbf{C_r}\in\mathbb{R}^M$ denotes the raw measured capacitances. $\mathbf{C_h}\in\mathbb{R}^M$ and $\mathbf{C_l}\in\mathbb{R}^M$ represent the raw measured capacitances when the sensing region is filled with high and low permittivity materials, respectively. $\mathbf{g}\in\mathbb{R}^N$ represents the normalized permittivity distribution; $\mathbf{S}\in\mathbb{R}^{M\times N}$ ($M<<N$) is the normalized sensitivity matrix. The inverse problem of 3D-ECT is to estimate $\mathbf{g}$ based on $\mathbf{C}$ and $\mathbf{S}$. The simplest estimation of $\mathbf{g}$ is to use linear back projection (LBP) [17]

$$\hat{\mathbf{g}} = \mathbf{S}^T\mathbf{C} \quad (3)$$

where $\hat{\mathbf{g}} \in\mathbb{R}^N$ is the estimation of $\mathbf{g}$. LBP has been widely applied in ECT, but it provides limited resolution and accuracy.

### B. Digital twin-assisted 3D-ECT

The flowchart depicted in Fig. 2 outlines the key components of our extended DT framework for 3D multiphase flow imaging. We first create a virtual representation of the 3D-ECT system, i.e., a 3D fluid-electrostatic field coupling model (3D-FECM). Through field coupling simulation, seamless data exchange is achieved between the virtual 3D-ECT sensor and the dynamic flow model. This enables real-time imaging and analysis of multiphase flows. Additionally, we develop a deep neural network, i.e., 3D deep back projection (3D-DBP), which takes LBP results as input and refines the LBP-reconstructed 3D images with a modified UNet [18]. We conducted large-scale virtual experiments that simulated various working conditions of a real multiphase flow facility. It allows us to synthesize a comprehensive 3D multiphase flow imaging dataset, comprising flow data and 3D-ECT measurements. The dataset will facilitate the training and evaluation of the 3D-DBP.

#### 1) 3D fluid-electrostatic field coupling model

The 3D-FECM incorporates a fluid field interface to generate gas-liquid two-phase flows and an electrostatic field interface to model the 3D-ECT sensor. By coupling the fluid and electrostatic fields, we can collect dynamic phase distributions and corresponding capacitance measurements from the virtual 3D-ECT sensor. To simulate the gas-liquid two-phase flows, we employ the laminar two-phase flow interface described by the incompressible Navier-Stokes equations [19]:

$$\rho\frac{\partial\mathbf{u}}{\partial t} + \rho(\mathbf{u}\cdot\nabla)\mathbf{u} = \nabla\cdot\left[-p\mathbf{I} + \mu\left(\nabla\mathbf{u} + \nabla\mathbf{u}^T\right)\right] + \mathbf{F} + \mathbf{F}_g \quad (4)$$

$$\nabla\cdot\mathbf{u} = 0 \quad (5)$$

where $\rho=\rho_l+(\rho_g-\rho_l)\phi$ represents the mixture density. $\rho_l$ and $\rho_g$ represent the densities of the liquid and gas, respectively; $\phi$ denotes the gas void fraction. $\mu=\mu_l+(\mu_g-\mu_l)\phi$ is the mixture dynamic viscosity. $\mu_l$ and $\mu_g$ represent the dynamic viscosities of the liquid and gas, respectively. $p$ denotes the pressure; $\mathbf{u}$ is the mixture velocity vector; $\mathbf{F}$ represents the volume force vector; $\mathbf{F}_g$ denotes the gravity force vector; $\mathbf{I}$ is the identity tensor. The level set method [20] is employed to track the boundary between the gas and liquid phases:

$$\frac{\partial\phi}{\partial t}+\mathbf{u}\cdot\nabla\phi=\gamma\nabla\cdot\left(\delta\nabla\phi-\phi(1-\phi)\frac{\nabla\phi}{|\nabla\phi|}\right) \tag{6}$$

where $\gamma$ represents the reinitialization parameter and is typically set to 1. $\delta$ denotes the interface thickness controlling parameter. In this work, we set $\delta$ to $h_{max}/2$, where $h_{max}$ is the maximum element size in the component. In the electrostatics interface, the electric potential in response to the permittivity can be expressed by [21]

$$-\nabla\cdot\left(\varepsilon_o\varepsilon\nabla\varphi\right)=0 \tag{7}$$

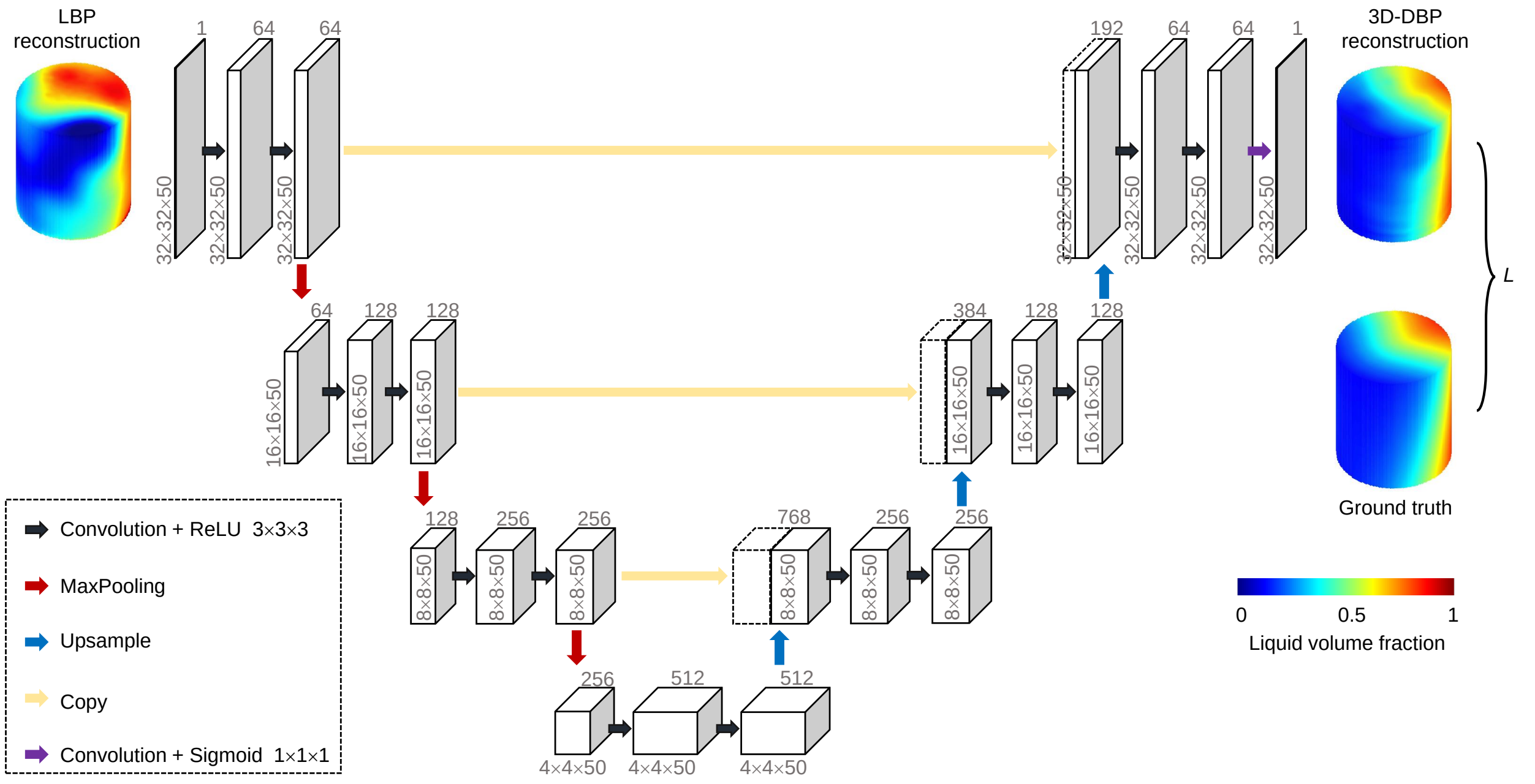

Fig. 3. The network architecture of 3D-DBP.

Here, $\varepsilon_o$ represents the vacuum permittivity.
To determine the equivalent permittivity of the mixture, we utilize the Wiener Upper Bound formula [22], given by

$$\varepsilon=\varepsilon_g\phi+\varepsilon_l(1-\phi) \tag{8}$$

where $\varepsilon_g$ and $\varepsilon_l$ denote the permittivity of gas and liquid phases, respectively.

#### 2) *3D deep back projection*

3D-DBP aims to learn the dynamic flow features in the virtual space and facilitate quantitative 3D multiphase flow imaging in the physical space. 3D-DBP involves two key steps. First, the capacitance measurements obtained from 3D-FECM are mapped into a coarse 3D flow distribution utilizing LBP. Subsequently, a modified 3D-UNet is employed to refine the coarse 3D images. Fig. 3 shows the encoder-decoder structure of the 3D-UNet, which mainly consists of a contracting path and a corresponding expanding path. The contracting path applies four convolutional blocks, each consisting of two 3D convolutional layers followed by ReLU activation. After each convolutional block, max pooling is applied to reduce the spatial dimensions of the feature maps while increasing their depth. The expanding path starts by upsampling the feature maps from the contracting path. The upsampled feature maps are then concatenated with the corresponding feature maps from the contracting path. Following concatenation, the resulting feature maps undergo a series of double convolution blocks. Each double convolution block consists of two consecutive 3D convolution layers, followed by ReLU activation. After passing through the double convolution blocks, the final feature maps are fed into the output layer.

## III. 3D Multiphase Flow Dataset Generation

### *A. Setup*

Fig. 4(a) presents the schematic diagram of the multiphase flow facility at the Multiphase Flow Engineering Laboratory of the Tsinghua International Graduate School. The facility comprises a multiphase flow separator, a gas storage tank, distinct sections dedicated to gas and liquid single-phase flows, a testing section, and a sophisticated control system. We developed a 3D-FECM (see Fig. 4(b)) using the 3D-ECT sensor as the virtual representation of the testing section of the multiphase flow facility (see the dashed line box in Fig. 4(a)). In the physical facility, gas and liquid single-phase flows are supplied and controlled separately, enabling the generation of gas–liquid flows with diverse volumetric concentrations. Similarly, within the 3D-FECM, dynamic gas and liquid flows are individually regulated to simulate a wide range of gas-liquid flow scenarios.

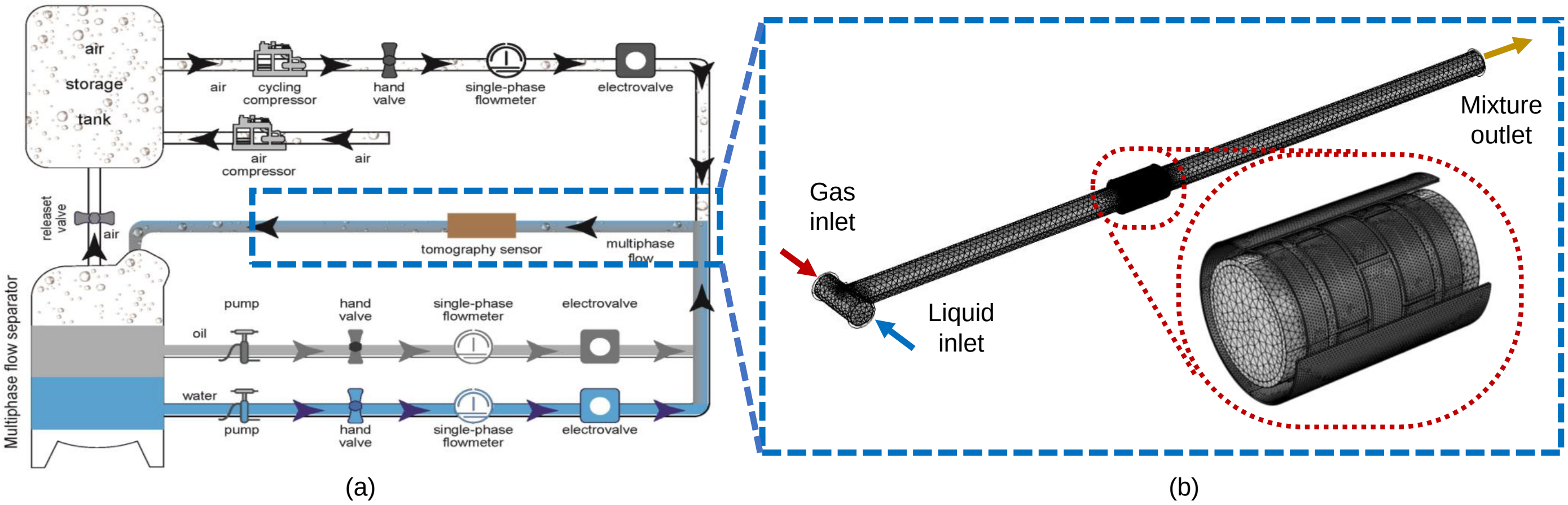

Fig. 4. Pilot-scale multiphase flow testing facility at the Multiphase Flow Engineering Laboratory of the Tsinghua International Graduate School. (a) Schematic illustration of the pilot-scale multiphase flow testing facility; (b) The model of the testing section of the multiphase flow facility.

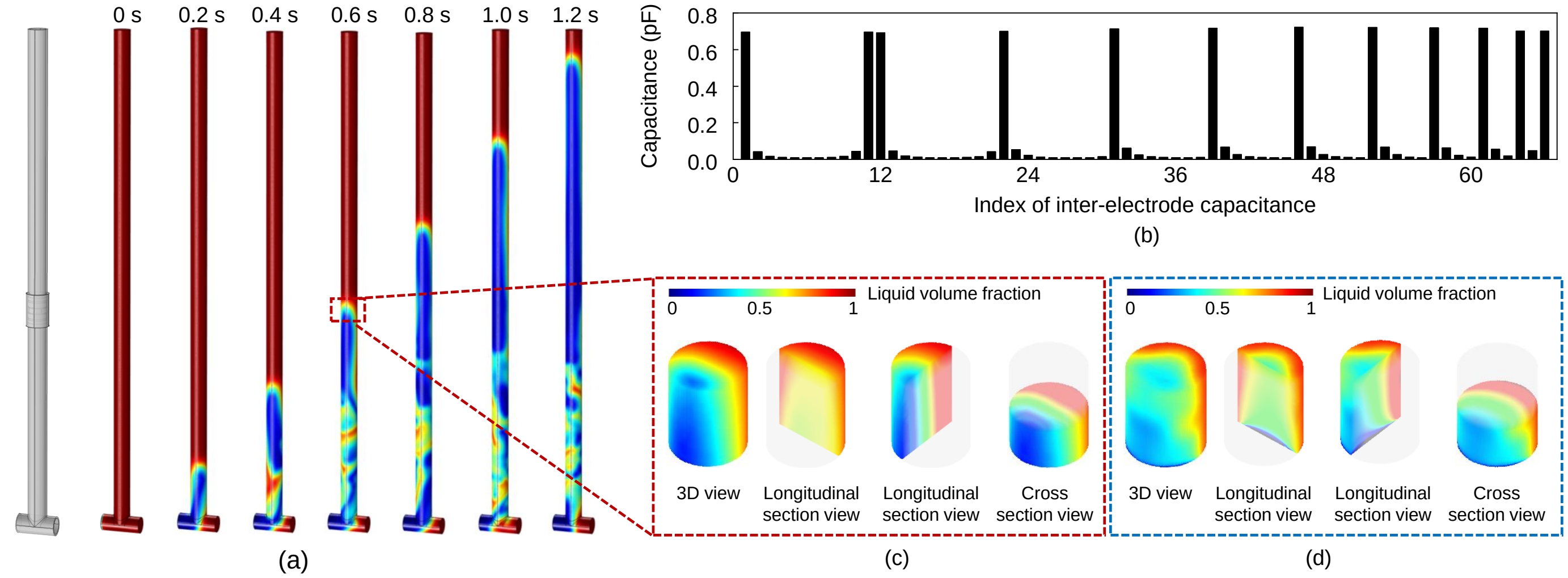

Fig. 5. Representative coupling simulation results generated by 3D-FECM. (a) Sequential gas-liquid flows generated by 3D-FECM; (b) Inter-electrode capacitance obtained by 3D-FECM; (c) 3D permittivity distribution in 3D-ECT's sensing region; (d) 3D images reconstructed by LBP.

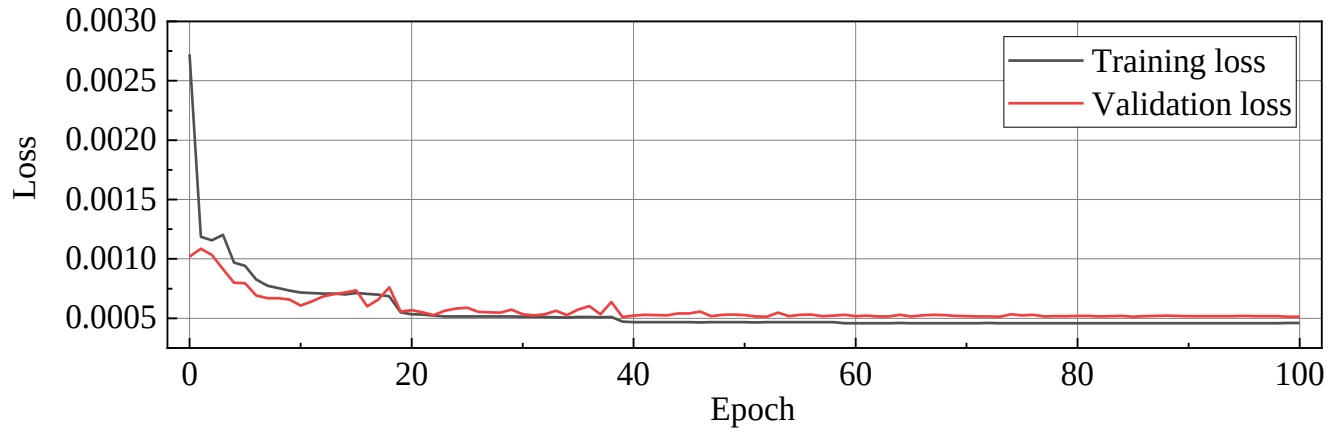

Fig. 6. Learning curve of the 3D-DBP.

Our previous work has demonstrated the accuracy and reliability of 3D-FECM [15], [16]. In this work, virtual experiments were conducted using parameter settings consistent with those in [16]. In the electrostatics interface, the relative permittivity for the pipe, gas and liquid was 2.6, 1, and 2.18, respectively. Inter-electrode capacitance data were measured. Regarding the fluid field interface, the gas phase represented air and the liquid phase was industrial white oil. The air was characterized by a density of 1.3 kg/m$^3$ and a dynamic viscosity of 1.81E-5 Pas, while the oil had a density of 879 kg/m$^3$ and a dynamic viscosity of 0.02 Pas. The inlet gas and liquid velocities ranged from 0.236 to 2.362 m/s and 0.071 to 0.708 m/s, respectively. Two initial conditions, i.e., one with the pipe initially filled with liquid and the other with gas, were implemented to encompass diverse volumetric concentrations.

## B. *Simulation results*

Fig. 5(a) presents sequential gas-liquid flows captured by the 3D-FECM at a 0.2 s interval. The inflow velocities of oil and gas are set at 0.709 m/s and 0.425m/s, respectively. Initially, the pipe was filled with liquid. With the injection of gas and liquid, gas-liquid flows gradually emerge and move towards the outlet. Through the couplings of fluid and electrostatic fields, the virtual 3D-ECT sensor manifested a specific electric potential distribution, enabling the acquisition of 66 independent inter-electrode capacitance data (see Figs. 5 (b) and (c)). Using LBP, we reconstructed the permittivity distribution within the 3D-ECT sensor under 50 dB signal-noise ratio (SNR) (see Fig. 5(d)). Although the LBP-reconstructed images exhibit similar permittivity variations to the ground truth for the given phantom, they suffer from noticeable artifacts and distortion with the structural similarity index (SSIM) of 0.9073, root mean square error (RMSE) of 0.0917, and peak signal-to-noise ratio (PSNR) of 20.2167 dB.

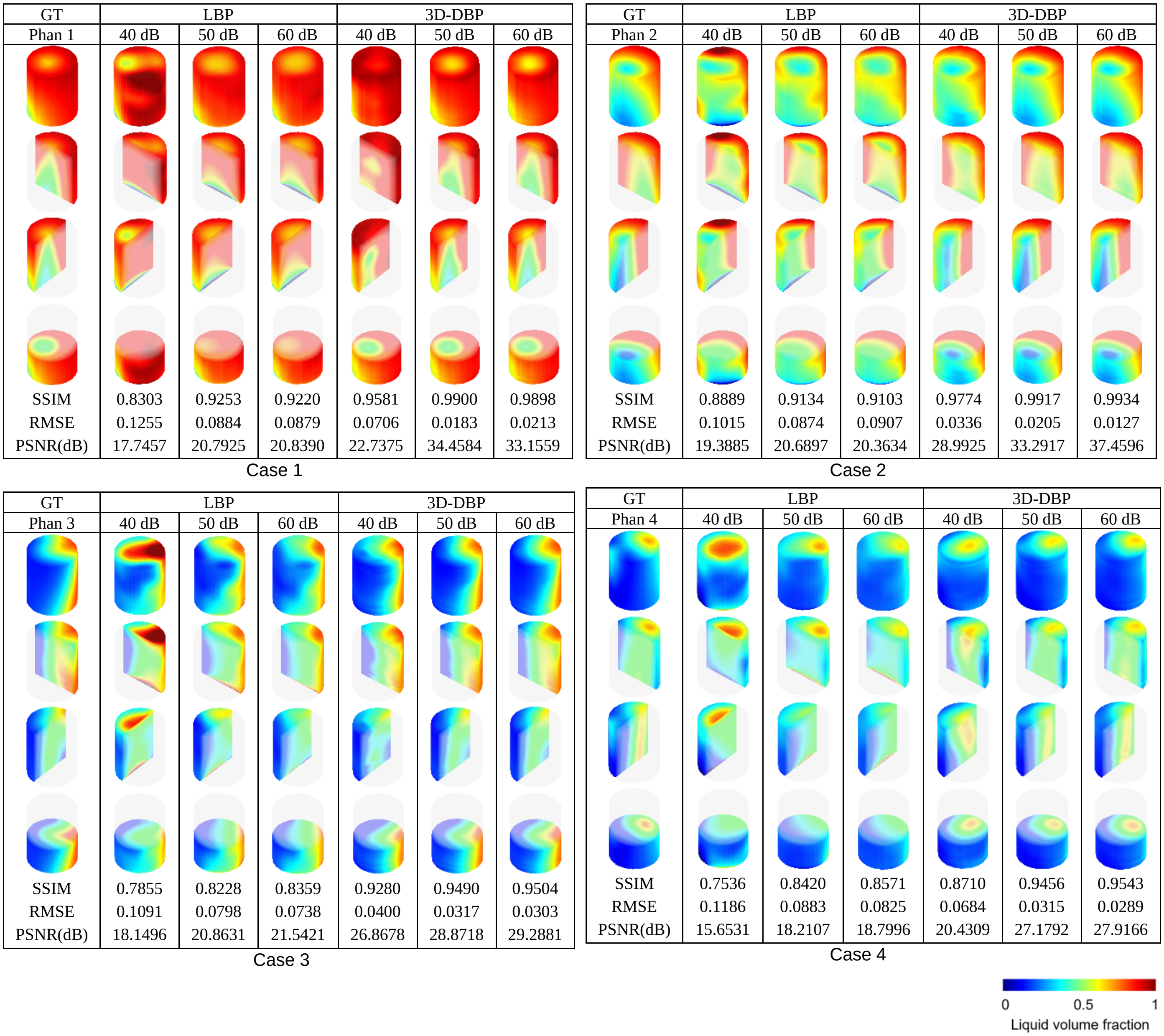

Fig. 7. Quantitative 3D imaging of virtual gas-liquid flows by LBP and 3D-DBP.

We conducted extensive virtual experiments to replicate the operational conditions of the actual multiphase flow testing facility. The inlet gas velocity varied from 0.236 to 2.362 m/s, while the inlet liquid velocity ranged from 0.071 to 0.708 m/s. In real-world multiphase flow measurements, the capacitance data obtained from our ECT system typically exhibit an SNR over 60 dB. To mimic the real ECT performance, we introduced three levels of additive noise (SNR 60, 50, and 40 dB) to the virtual capacitance data. In total, we simulated 60 working conditions, resulting in a comprehensive dataset of 12,362 virtual samples that encompass 3D ECT measurements and phase distributions.

## IV. Results and Discussion

### *A. Network Training*

To train the 3D-DBP network, an Adam optimizer with a learning rate of 1e-3 and weight decay of 5e-6 was used for optimization. To facilitate faster convergence and improve accuracy, a step decay strategy was employed, wherein the learning rate was reduced by a factor of 0.2 every 20 epochs. The training was implemented using NVIDIA P5000 GPUs. A batch size of 16 samples was utilized and the network underwent 100 iterations over the training dataset. Fig. 6 shows the learning curves, displaying the training and validation loss of the models. The model exhibiting the minimum validation loss was chosen as the final trained model.

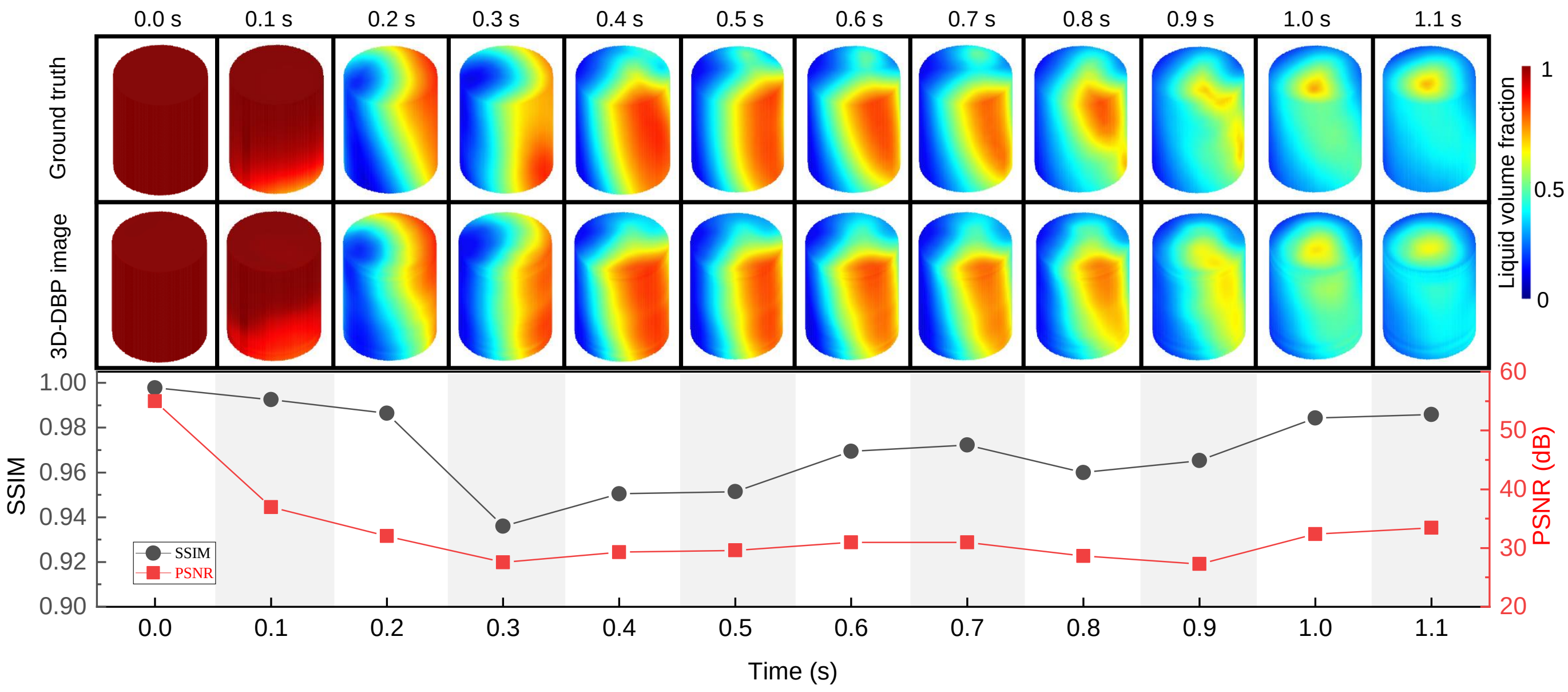

Fig. 8. Continuous reconstruction results of 3D-DBP for virtual gas-liquid flows with 60 dB SNR.

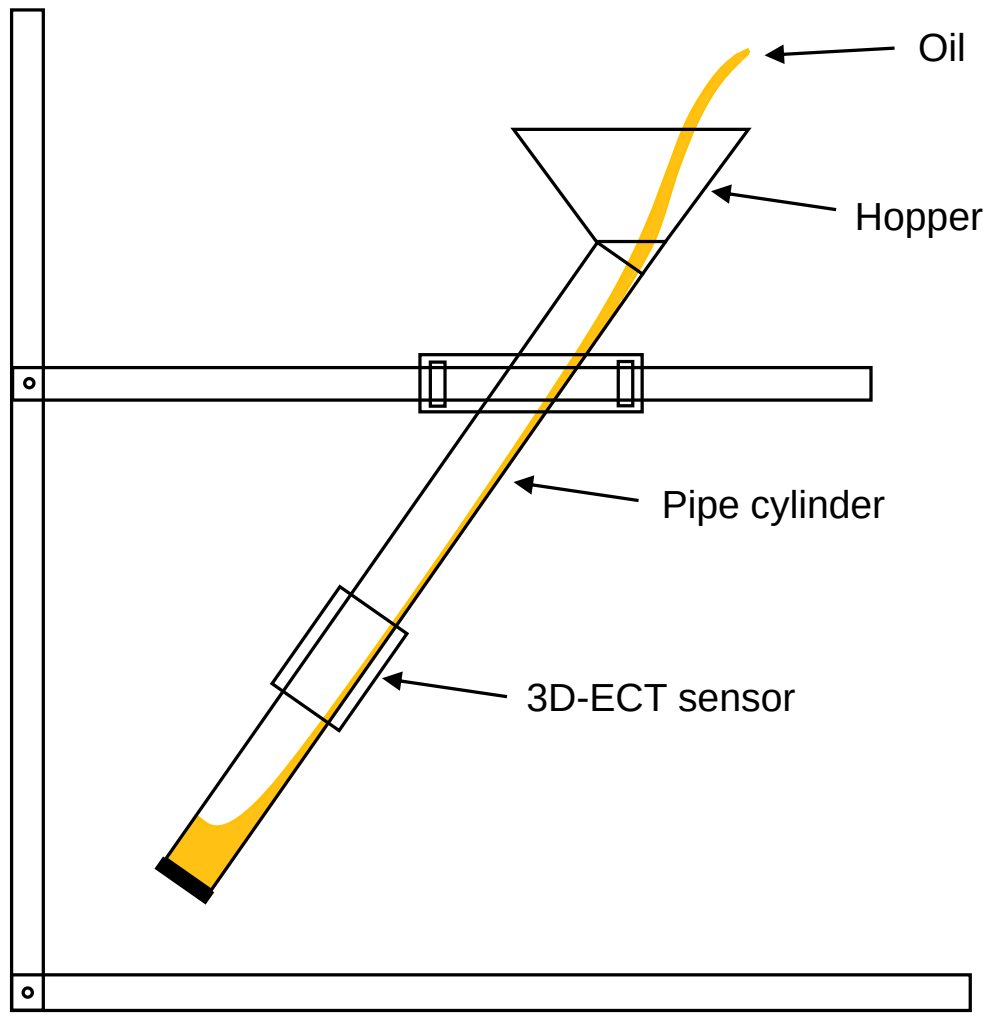

Fig. 9. Schematic of the gravity-driven oil-gas two-phase flow rig

## *B. Results for virtual gas-liquid flow imaging*

Fig. 7 presents the reconstructed images by LBP and 3D-DBP, and corresponding quantitative evaluations including SSIM, RMSE and PSNR, for four representative phantoms. The image reconstruction process incorporated three levels of additive noise SNR, ranging from 40 to 60 dB. Both LBP and 3D-DBP results exhibit similar permittivity variations to the ground truth for the given phantoms. It is evident that the reconstructed images using LBP show more significant distortion and deterioration, particularly near the boundary of the imaging region, in contrast to the 3D-DBP results. The quantitative evaluation further validates the relatively inferior performance of LBP, as evidenced by its lowest SSIM value of 0.7536, largest RMSE of 0.1255, and lowest PSNR of 15.6531 dB. Conversely, the 3D-DBP reconstructed images exhibit higher SSIM values (above 0.8710), smaller RMSE (below 0.0706) and higher PNSR (above 20.4309 dB). This indicates superior imaging quality and more accurate permittivity estimations achieved by 3D-DBP. Consequently, the 3D-DBP model effectively learns the flow features in the virtual space.

Additionally, we observe that as the SNR increases from 40 to 60 dB, both LBP and 3D-DBP results show increased SSIMs and PNSRs, and decreased RMSEs. However, the increase in SNR results in smaller improvements for the 3D-DBP results, with an increase of 0.0317, 0.016, 0.0224 and 0.0833 in SSIM for Phantoms 1-4, respectively. In contrast, the improvements for the LBP results are more significant, with SSIMs increasing by 0.0917, 0.0214, 0.0504, and 0.1035. This indicates that 3D-DBP exhibits excellent noise reduction performance.

Furthermore, we conducted continuous measurements of virtual gas-liquid flow at intervals of 0.1 seconds to further evaluate the performance of 3D-DBP. Fig. 8 shows the image reconstruction results of 3D-DBP for a series of sequential virtual gas-liquid flows with 60 dB SNR. The reconstructed images closely resemble the ground truth, with SSIM and PNSR exceeding 0.936 and 27.275 dB, respectively, indicating that the trained 3D-DBP can achieve accurate imaging of complex dynamic gas-liquid flows in the virtual space.

## *C. Results for physical gas-liquid flow imaging*

We conducted gas-liquid dynamic flow imaging experiments to assess the performance of the DT framework using a gravity-driven oil-gas two-phase flow rig. The schematic diagram in Fig. 9 illustrates the setup of the rig, which consists of a pipe cylinder inclined at an angle of approximately 60° relative to the horizontal plane. The experiment involved carefully pouring oil into the pipe cylinder from a hopper, allowing it to flow downward along the inner wall and pass through the 3D-ECT sensor. As the oil injection continued, the accumulated oil gradually rose within the cylinder, resulting in an increasing oil level that eventually submerged the sensing area of the 3D-ECT sensor.

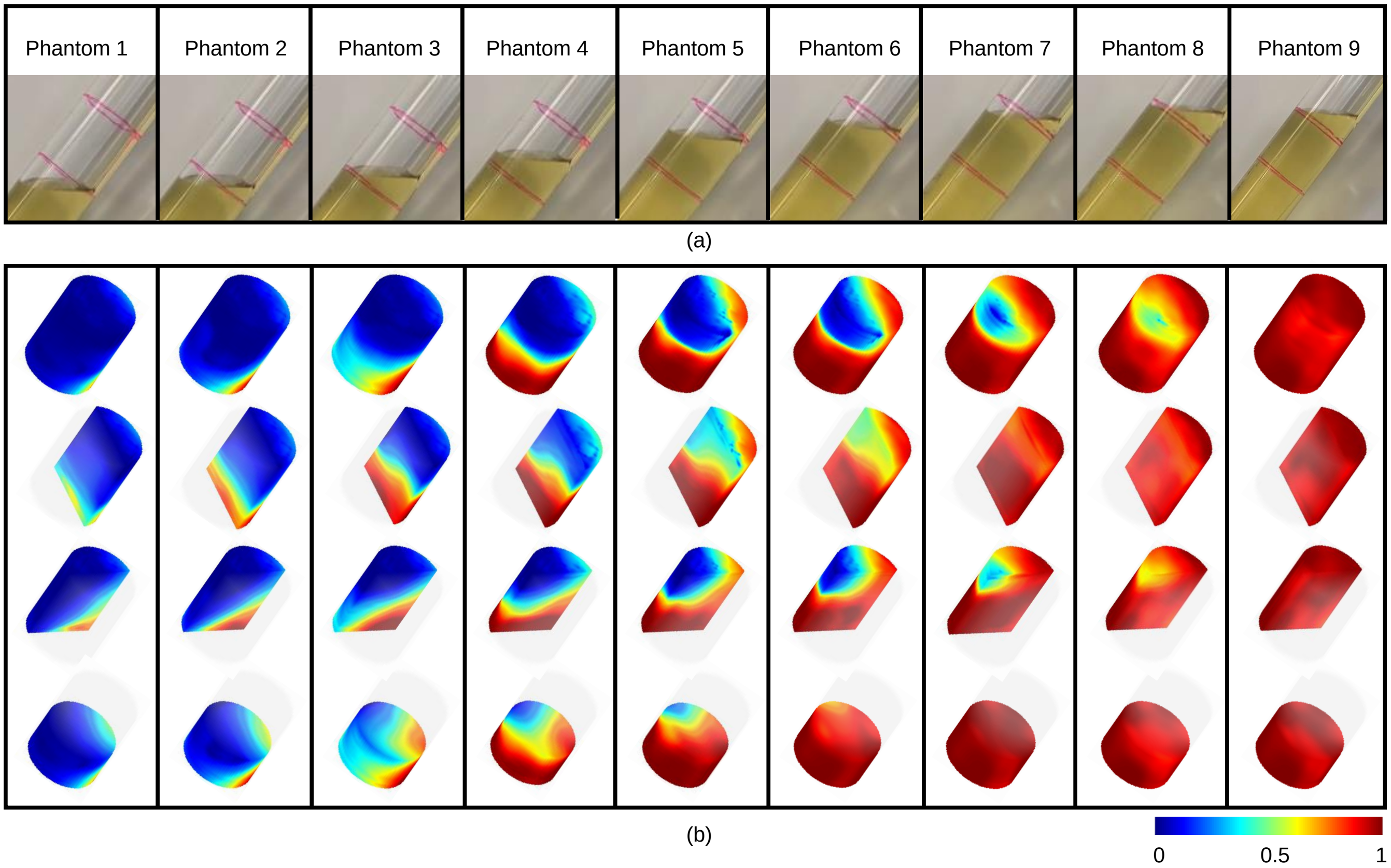

Fig. 10. 3D imaging of gas-liquid flows in the gravity-driven oil-gas two-phase flow rig. (a) Flow profiles captured by the camera; (b) 3D images obtained from our DT framework. Two red lines define the sensing region.

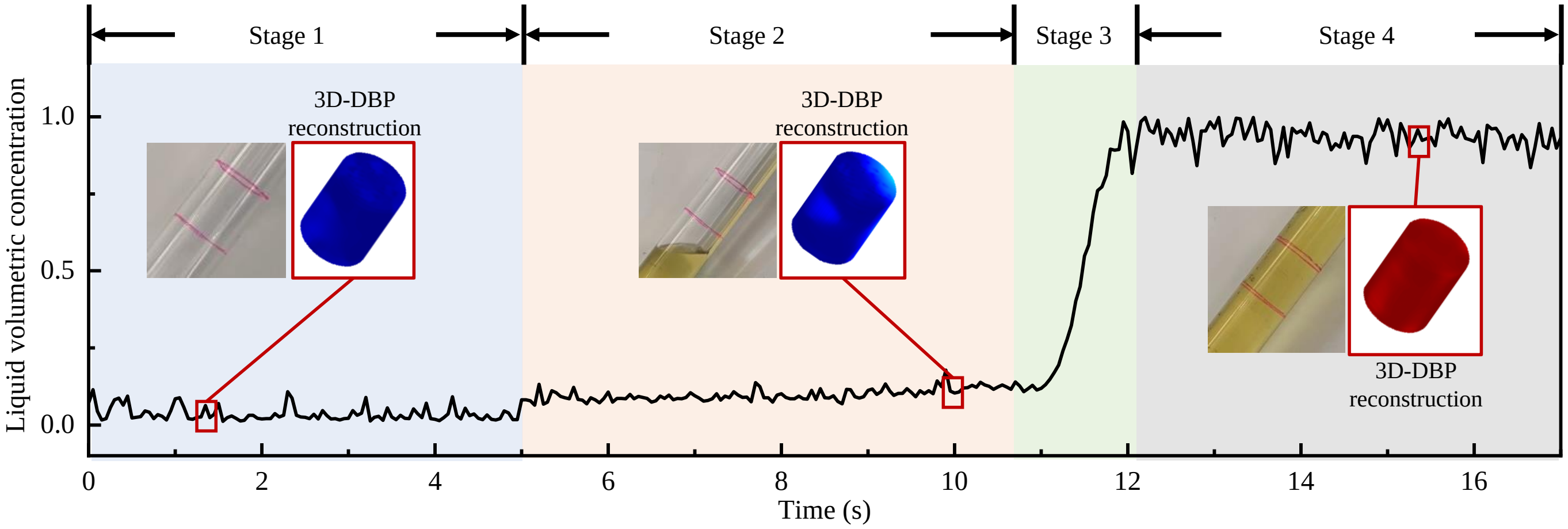

Fig. 11. Continuous measurements of the liquid volumetric concentration (LVC) of the dynamic gas-liquid flow during the oil pouring process.

The imaging experiment capturing the complete oil perfusion process can be divided into four stages, designated as stages 1-4. In stage 1, the pipeline is initially filled with air, and the oil injection has not yet commenced. Stage 2 signifies the commencement of oil filling in the pipeline, but the oil layer has not yet reached the sensing area of the 3D-ECT sensor. In stage 3, the oil layer enters the sensing range of the sensor. Finally, in stage 4, the sensing range of the 3D-ECT sensor is filled with oil.

To visually illustrate the oil pouring process, we captured examples of oil infusion by extracting representative frames at about 0.1-second intervals in stage 3, as shown in Fig. 10(a). The red circle on the pipe indicates the 3D-ECT sensor's sensing area. Fig. 10(b) presents the continuous imaging results obtained using the DT framework. Notably, we observed a close resemblance between the trends of phase distribution changes in the tomographic image and the photos during the oil perfusion process. At each time step, the oil and gas phase distribution captured by the tomographic image closely matched that depicted in the photos.

Fig. 11 shows the continuous measurements of the liquid volumetric concentration (LVC) of the dynamic gas-liquid flow during the oil pouring process. It is evident that during stage 1, the LVC measurements are approximately 0, indicating the absence of liquid in the sensing area. As it proceeds to stage 4, the measured LVC value exhibits fluctuations around 1, indicating that the sensing area of the 3D-ECT sensor is completely filled with oil. The standard deviation of the LVC measurements is 0.023 for stage 1 and 0.036 for stage 4, demonstrating the superior stability of the DT framework.

## V. Conclusion

This paper presents a digital twin (DT) framework for addressing the challenges associated with 3D electrical capacitance tomography (3D-ECT) in imaging multiphase flows. The DT framework integrates the 3D fluid-electrostatic field coupling model (3D-FECM) and a deep neural network called 3D deep back projection (3D-DBP) to enhance the visualization and efficiency of multiphase flow imaging systems. The combination of the 3D-FECM and 3D-DBP enhanced the resolution, accuracy, and reliability of reconstructed images, surpassing the traditional method. The comparison with ground truth and evaluation metrics validated the superiority of the 3D-DBP network in terms of resolution, accuracy, and structural similarity. This research contributes to advancing 3D-ECT for imaging multiphase flows in various industries, offering a robust inversion algorithm, a virtual evaluation platform, and a promising approach for process understanding and optimization.